\documentclass[graybox]{svmult}

\usepackage{mathptmx}       % selects Times Roman as basic font
\usepackage{helvet}         % selects Helvetica as sans-serif font
\usepackage{courier}        % selects Courier as typewriter font
\usepackage{type1cm}        % activate if the above 3 fonts are
\usepackage{makeidx}         % allows index generation
\usepackage{graphicx}        % standard LaTeX graphics tool
\usepackage{multicol}        % used for the two-column index
\usepackage[bottom]{footmisc}% places footnotes at page bottom

\usepackage{bbold}
\makeindex             % used for the subject index
\begin{document}

%10-15 pages

\title*{Aftershock prediction for high-frequency financial markets' dynamics}
%\titlerunning{Aftershock prediction in high-frequency financial data}

\author{Fulvio Baldovin, Francesco Camana, Michele Caraglio, Attilio L. Stella, Marco Zamparo}
\authorrunning{F. Baldovin, F. Camana, M. Caraglio, A.L. Stella, M. Zamparo}

\institute{
Fulvio Baldovin \at Dipartimento di Fisica, Sezione INFN and Sezione CNISM
Universit\`a di Padova, Via Marzolo 8, I-35131 Padova, Italy \email{baldovin@pd.infn.it}
\and
Francesco Camana \at Dipartimento di Fisica 
Universit\`a di Padova, Via Marzolo 8, I-35131 Padova, Italy \email{camana@pd.infn.it}
\and
Michele Caraglio \at Dipartimento di Fisica 
Universit\`a di Padova, Via Marzolo 8, I-35131 Padova, Italy \email{caraglio@pd.infn.it}
\and 
Attilio L. Stella \at Dipartimento di Fisica, Sezione INFN and Sezione
CNISM
Universit\`a di Padova, Via Marzolo 8, I-35131 Padova, Italy \email{stella@pd.infn.it}
\and 
Marco Zamparo \at Dipartimento di Fisica, Sezione INFN and Sezione CNISM
Universit\`a di Padova, Via Marzolo 8, I-35131 Padova, Italy;\\
HuGeF, Via Nizza 52, 10126 Torino, Italy 
\email{marco.zamparo@hugef-torino.org} 
}
\maketitle

\abstract*{
  The occurrence of aftershocks following a major financial crash
  manifests the critical dynamical response of financial markets. 
  Aftershocks put additional stress on markets, with conceivable
  dramatic consequences. Such a phenomenon has been shown to be
  common to most financial assets, both at high and low frequency. 
  Its present-day description relies on an empirical characterization
  proposed by Omori at the end of 1800 for seismic earthquakes. We
  point out the limited predictive power in this phenomenological
  approach and present a stochastic model, based on the scaling
  symmetry of financial assets, which is potentially capable to
  predict aftershocks occurrence, given the main shock
  magnitude. Comparisons with S\&P high-frequency data confirm this
  predictive potential.
}

\abstract{
  The occurrence of aftershocks following a major financial crash
  manifests the critical dynamical response of financial markets. 
  Aftershocks put additional stress on markets, with conceivable
  dramatic consequences. Such a phenomenon has been shown to be
  common to most financial assets, both at high and low frequency. 
  Its present-day description relies on an empirical characterization
  proposed by Omori at the end of 1800 for seismic earthquakes. We
  point out the limited predictive power in this phenomenological
  approach and present a stochastic model, based on the scaling
  symmetry of financial assets, which is potentially capable to
  predict aftershocks occurrence, given the main shock
  magnitude. Comparisons with S\&P high-frequency data confirm this
  predictive potential.
}

\section{Introduction}
\label{sec_introduction}
It is not uncommon for financial indexes or asset prices to experience 
exceptionally large negative or positive returns which trigger periods 
of high volatility, the case of abnormal negative returns 
corresponding to market crashes. An understanding of the dynamical
response of the market to a {\it main shock} is of great interest
because it may help, e.g., in the definition of emergency plans for 
financial crises, or for risk management.

There is a clear analogy between the behavior of volatility after
a main financial shock and that of the seismic activity after an
earthquake of exceptional magnitude in geophysics \cite{scholz}. 
Omori \cite{omori}, with a subsequent modification by Utsu \cite{utsu}, established 
an important empirical law describing the frequency of 
occurrence of seismic events above a given threshold after a main 
earthquake.   
The characterizing feature of this law is the decay as a power of time, $t$,
of the rate of occurrence of aftershocks above the threshold, indicating the 
absence of a characteristic time scale in the manifestly non-stationary 
Omori regime. 
More precisely, according to Omori the number, $n(t)$, of aftershocks per unit time 
above a given threshold $\sigma_a$ is given by 
\begin{equation}
n(t)=K\;(t+\tau)^{-p},
\end{equation}
where $K,\tau,p$ depend on the aftershock threshold
$\sigma_a$, and also on the specific magnitude of the main shock
earthquake. 
Equivalently, the Omori law can be expressed in an integral form as 
\begin{equation}
N(t)=\frac{K}{1-p}\;\left[(t+\tau)^{1-p}-\tau^{1-p}\right]
\label{eq_omori_cumulative}
\end{equation}
if $p\neq1$, or $N(t)=K\;\ln(t/\tau+1)$ if $p=1$,
where $N(t)$ is the cumulative number of aftershocks up to time $t$
after the main shock. 
Lillo and Mantegna \cite{lillo} were the first to verify the validity 
of an analog of the Omori law for the volatility in Finance after a 
main crash. They also showed \cite{lillo_1} that standard
dynamical models of index evolution, like GARCH, are not adequate to 
reproduce financial Omori-like regimes.
Several studies \cite{lillo,lillo_1,selcuk,selcuk_1,weber,mu,petersen} verified the
presence of Omori regimes under various market conditions, triggered
by financial crashes \cite{lillo,lillo_1,selcuk,selcuk_1,weber}, by
volatility shocks \cite{mu}, and even by U.S. Federal Open Market
Committee meetings \cite{petersen}.   
In particular, the Omori law in finance
has been upgraded to a more general characterization of 
market dynamics by Weber et al. \cite{weber}, who pointed out that this 
law holds on a wide range of time scales, with aftercrashes of a 
main shock playing the role of main crashes for even smaller 
aftercrashes, etc.

The above mentioned studies make clear the connection between financial
Omori processes and long-range dependence of the volatility. They also show
that a modulating, time dependent scale for the returns must be considered in 
order to account for the manifest non-stationarity of the Omori process. At 
the same time, they emphasize the limits in the predictive value of the Omori 
law. For example, the parameters $K$ and $\tau$ need to be adjusted for each
aftershock threshold considered (See below). This holds also for the exponent
$p$ of the power law decay, which should be expected to be the most robust
parameter. In addition, there is no idea of how the parameters could be
linked to the magnitude of the main shock. 
These limits reflect a lack of adequate modeling for the
dynamics of financial indexes, especially in regimes like those
covered by the Omori law. In recent contributions 
\cite{baldovin,baldovin_1}, some of the
present authors have proposed a model for the dynamics at high frequency
of exchange rates or stock market indexes, which takes into account
most of the relevant stylized facts. Among them, the martingale character 
of index evolution, the manifest non-stationarity of volatility detected in 
well defined daily windows of trading activity, the anomalous scaling properties of 
the aggregate return probability density function (PDF) 
in the same windows, and the strong time autocorrelation 
of the elementary absolute return. This model for high-frequency data,
which applies more general ideas
about the time evolution of financial indexes
\cite{baldovin_0,stella,challet}, 
has also been tested \cite{baldovin} by comparing 
its predictions with the statistics of ensembles of daily histories
all supposed to reproduce the same underlying stochastic 
process.
It has been also shown \cite{baldovin_1}
that some arbitrage opportunities revealed by the model 
could be successfully exploited by appropriate trading strategies.

In the present contribution we address the problem of describing with 
such a model the Omori processes which may be detected within these 
daily windows. 
Our goal is to show that, after proper calibration,
this model allows the 
{\it prediction} of the aftershock rate within an Omori regime, 
given the value of the main shock magnitude.  
Indeed, we provide analytical 
expressions for the rate of financial aftershocks with explicit dependence 
on the magnitude of the main shock and on the aftershocks threshold. 
By comparing our predictions with high frequency data from the 
S\&P 500 index we show that these quantities are sufficient to  
determine the Omori response without further fitting parameters.
Our success is partly due to the fact that we are able to identify
the Omori processes within a context for which non-stationarity is
well established \cite{bassler} and amenable to modeling
\cite{baldovin,baldovin_1}. 
In an interday context, the question of the
applicability of the models of Ref. \cite{baldovin_0} to Omori regimes
has already
been raised in Ref. \cite{challet}. 

This note is organized as follows. In the next Section we briefly
recall the model of Refs. \cite{baldovin,baldovin_1} 
and present the procedure of calibration.
In the third Section we discuss the selection of Omori-like processes
from our database and show how our model can be used to
analytically describe these processes. 
In the fourth Section we compare the results 
of the properly calibrated model with the statistical records at our 
disposal for the S\&P 500 index. The last Section is 
devoted to general discussion and conclusions.

\section{Model calibration}
\label{sec_calibration}
Let us consider the successive (log-)returns over ten minutes intervals of
the S\&P 500 index $S(t)$ for daily windows from 9.40 a.m.,
Chicago time, to 1.00 p.m.:
\begin{equation}
R_t\equiv \ln S(t+1)-\ln S(t),\quad t=0,1,\ldots,19,
\end{equation} 
where the time is measured in ten-minute units and we have set $t=0$
at 9.40 a.m.\footnote{
In order to keep contact with ordinary notations for the Omori law, in
this paper we change slightly our usual conventions by shifting 
the origin of time by one
unit with respect to, e.g., Refs. \cite{baldovin,baldovin_1}.
}.  
A statistics made over the ensemble of 6283 available daily histories from 
1985 to 2010 shows \cite{baldovin_1} 
that a stochastic process supposed to generate the 
successive returns $R_t$ 
in a generic history of the ensemble is consistent 
with the following joint PDF:
\begin{equation}
p_{R_0,R_1,\ldots,R_t}(r_0,r_1, \dots, r_t)=\int_0^\infty d\sigma \;\rho(\sigma)
\;\prod_{i=0}^{t}\frac{\exp\left(-{{r_i^2}\over{2\sigma^2
      a_i^2}}\right)}{\sqrt{2\pi\sigma^2 a_i^2}},
\label{eq_joint} 
\end{equation}
where 
\begin{equation}
a_i=[(i+1)^{2D}-i^{2D}]^{1/2}
\end{equation}
with $D\geq0$, $i=0,1,\ldots,19$, and $\rho(\sigma)\geq0$ with
\begin{equation}
\int_0^\infty d\sigma \;\rho(\sigma)=1.
\end{equation}
This joint PDF is a convex combination of 
products of Gaussian PDF's for each individual return. The PDF $\rho(\sigma)$
weights this combination and introduces a nontrivial dependence of the
returns from the preceding ones. 
For $D\neq 1/2$, the coefficients $a_i$
make the process increments
non-stationary, and modulated by the 
exponent $D$. 

\begin{figure}[t]
%\sidecaption[t]
\begin{center}
\includegraphics[scale=.5]{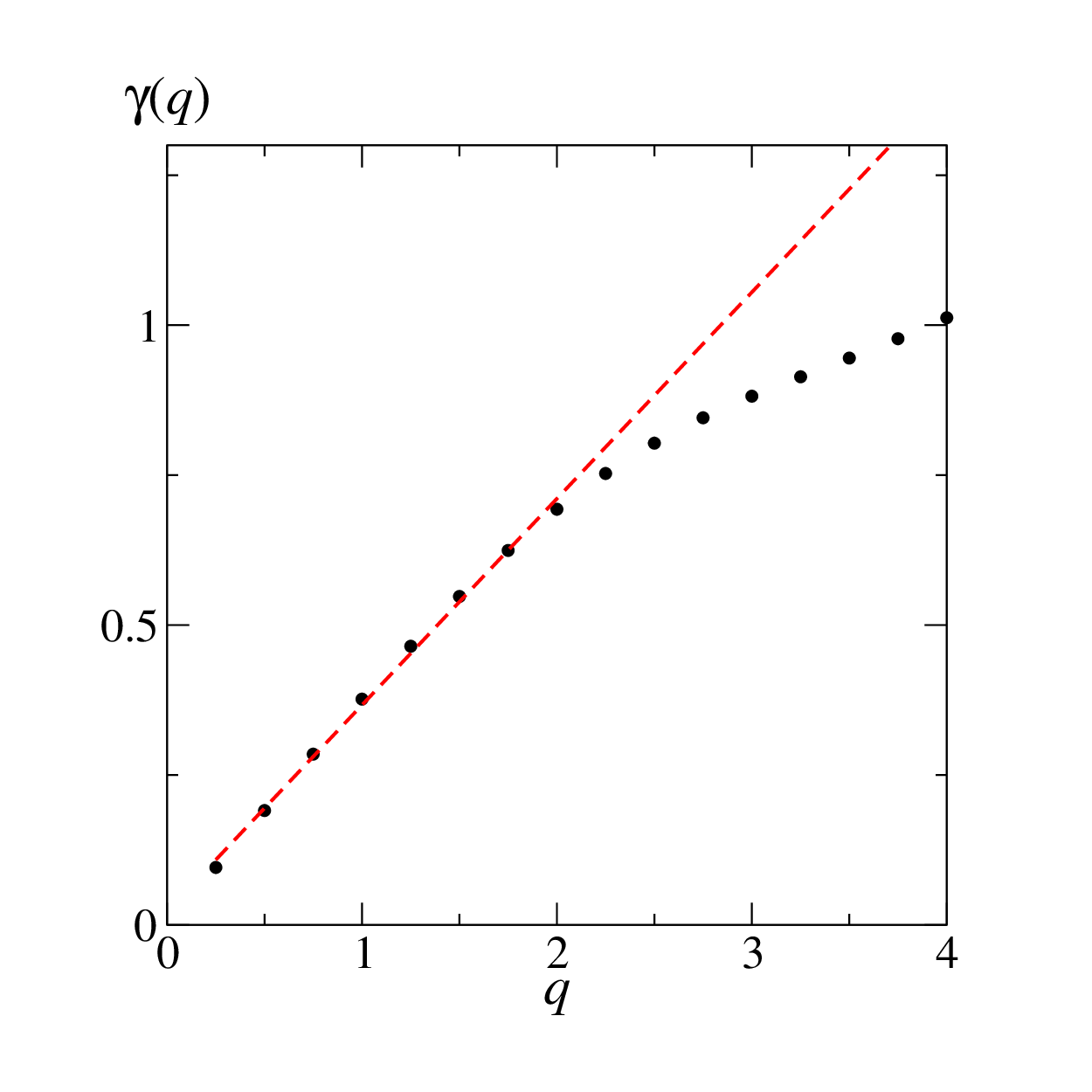}
\end{center}
\caption{
Calibration of the scaling exponent $D$. The empirical values for
$\gamma(q)$ are reported using Eq. (\ref{eq_scaling}) as an ansatz (points). 
A linear regression for
$0<q\leq 2$ gives $\gamma(q)=q\;D$ with $D\simeq 0.35$ (dashed line).
}\label{fig_scaling} 
\end{figure}

The calibration of the model can be done by direct comparison of 
its predictions with the main features of the PDF's of the 10-minute
returns $R_i$'s,
or, alternatively, 
with those of the aggregate returns $\sum_{i=0}^t R_i$
\cite{baldovin,baldovin_1}. 
Here we follow the second option. 
Since the model predicts for the PDF of the
aggregate return $\sum_{i=0}^t R_i$ satisfaction of an anomalous scaling 
of the form
\begin{equation}
p_{\sum_{i=0}^t R_i}(r)=\frac{1}{(t+1)^D}\;g\left(\frac{r}{(t+1)^D}\right),
\label{eq_aggregate_scaling}
\end{equation}
where the scaling function $g$ is expressed as
\begin{equation}
g(r)=\int_0^\infty d\sigma\; \rho(\sigma) 
\frac{
\exp\left(-\frac{r^2}{2\sigma^2}\right)
}{
\sqrt{2\pi\sigma^2}
},
\end{equation}
one can determine $D$ through a fitting of the power law $t$-dependence of the  
moments of $p_{\sum_{i=0}^t R_i}$. 
Indeed, for $q\in\mathbb R$, according to Eq. (\ref{eq_aggregate_scaling})
\begin{equation}
\mathbb E\left[\left|\sum_{i=0}^t R_i\right|^q\right]
=\mathbb E\left[\left|R_0\right|^q\right]\;t^{\gamma(q)}
\label{eq_scaling}
\end{equation}
with $\gamma(q)=q\;D$, and provided that the moment
$\mathbb E\left[\left|R_0\right|^q\right]$ exists. 
In Fig. \ref{fig_scaling} we report the empirical values for
$\gamma(q)$, using Eq. (\ref{eq_scaling}) as an ansatz. 
To calibrate $D$, we make a linear data regression for $q\leq2$,
since for higher moments a multiscaling behavior \cite{wang} is detected 
(See Fig. \ref{fig_scaling}). The result is $D\simeq0.35$.

\begin{figure}[t]
%\sidecaption[t]
\begin{center}
\includegraphics[scale=.5]{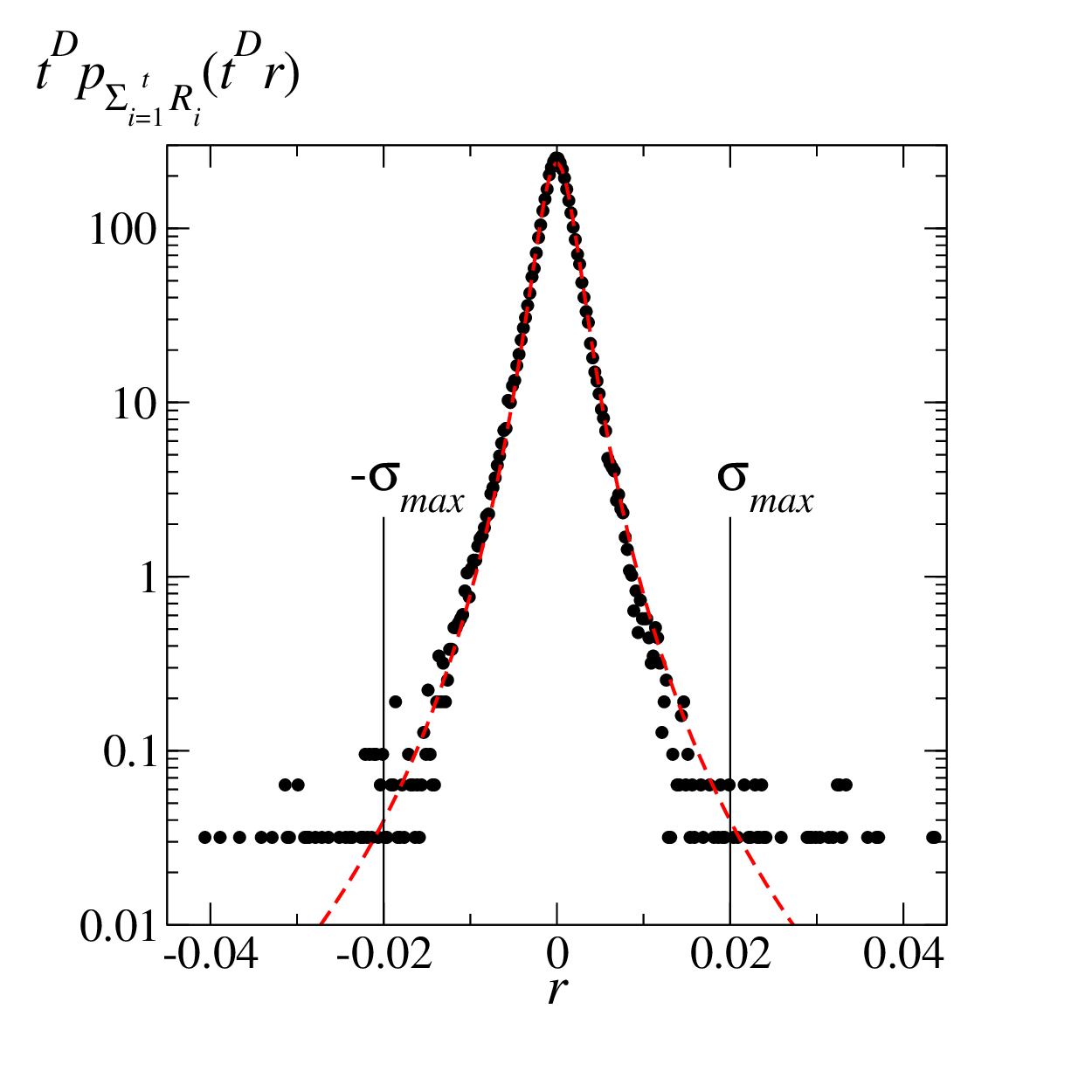}
\end{center}
\caption{
Calibration of the parameters $\alpha$ and $\beta$. 
The empirical PDF's for $\sum_{i=0}^t R_i$ at various $t$ are rescaled according
to Eq. (\ref{eq_aggregate_scaling}) with the previously calibrated
$D=0.35$ (points). The parameters $\alpha$ and $\beta$ are then
fitted using Eq. (\ref{eq_student}) with $t=0$ (dashed line), 
yielding the values
$\alpha=3.5$ and $\beta=2.9\cdot10^{-3}$.
An upper bound to the empirical analysis is posed at the
$\sigma_{max}=0.02$ for the 10-minute volatility. 
}\label{fig_scaling_function} 
\end{figure}

A particularly simple expression for the joint PDF
$p_{R_0,R_1,\ldots,R_t}$ is achieved if the integration on $\sigma$
can be worked out explicitly in Eq. (\ref{eq_joint}). 
This is indeed the case if we choose an
inverse-gamma distribution for $\sigma^2$
\cite{challet}. Equivalently, we may set
\begin{equation}
\rho(\sigma)=
\frac{2^{1-\frac{\alpha}{2}}\;\beta^\alpha}
{\Gamma\left(\frac{\alpha}{2}\right)\;\sigma^{\alpha+1}}
\;\exp\left(-\frac{\beta^2}{2\sigma^2}\right),
\label{eq_rho}
\end{equation}
where the exponent $\alpha$ determines the long-range behavior of $g$
according to $g(r)\sim 1/r^{\alpha+1}$ for $|r|\gg1$, and $\beta$ is a
scale parameter determining the distribution width.
Performing the integration on $\sigma$ in Eq. (\ref{eq_joint}) we
obtain a multi-variate Student PDF:
\begin{equation}
p_{R_0,R_1,\ldots,R_t}(r_0,r_1, \dots, r_t)
=\frac{
\beta^\alpha\;\Gamma\left(\frac{\alpha+t+1}{2}\right)
}{
\pi^{\frac{t+1}{2}}\;\Gamma\left(\frac{\alpha}{2}\right)
}
\;\left(
\beta^2+\frac{r_0^2}{a_0^2}+\frac{r_1^2}{a_1^2}+\cdots+\frac{r_t^2}{a_t^2}
\right)^{-\frac{\alpha+t+1}{2}}.
\label{eq_student}
\end{equation}
As we will show in the following, an explicit form for
$p_{R_0,R_1,\ldots,R_t}$ enables us to obtain a simple analytic
expression for $N(t)$.  
Unlike in previous papers \cite{baldovin,baldovin_1}, we
thus choose here the functional form in Eq. (\ref{eq_rho}) for $\rho$.
Besides $D$, the other parameters of the
model, $\alpha$ and $\beta$, are calibrated 
by first data-collapsing the empirical PDF's for 
$\sum_{i=0}^t R_i$ according to Eq. (\ref{eq_aggregate_scaling}) with $D=0.35$, 
and then by fitting $\alpha$ and $\beta$ on this data-collapse using 
Eq. (\ref{eq_student}) with $t=0$. The result is given in 
Fig. \ref{fig_scaling_function}.
In summary, the result of the calibration procedure is the triple
$(\alpha,\beta,D)=(3.5,2.9\cdot10^{-3},0.35)$.

The ensemble of histories at our disposal is relatively poor. 
This implies, as can be appreciated in
Fig. \ref{fig_scaling_function}, that some rare 
events fall significantly out of the scaling function, since a much
larger number of histories would be needed to correctly characterize
their frequency of occurrence. 
The multiscaling behavior shown in Fig. \ref{fig_scaling} could be at
least partly related to this effect.
The Omori events are precisely related to extreme events. 
In order to obtain a reliable statistics of the aftershocks, we  
impose thus an upper bound $\sigma_{max}$ to the absolute value of
the returns $R_i$'s included in our empirical analysis (See
Fig. \ref{fig_scaling_function}).   
Once done this, 
the overall agreement of the empirical data with the various  
model predictions gives a convincing validation of the model itself
(See also \cite{baldovin_1}). 
Still, the agreement shown in what follows with respect to the Omori
processes must be intended as a first important result, which 
calls for more extensive analysis also in terms of the calibration
procedure.

\section{Aftershock prediction}
As already mentioned above, in the present analysis we are going to
identify and select Omori processes, which are manifestations 
of non-stationarity, within a process which manifestly turns out to be with non-stationary
returns in its ensemble of daily realizations. This is a simplification
which marks an important difference with respect 
to the problem of modeling the Omori regimes revealed in Refs. 
\cite{lillo,lillo_1,selcuk,selcuk_1,weber,mu,petersen}, where they were  
extracted from single time series expected to be globally stationary on long time 
scales. In the perspective of our approach here, dealing with a process 
which is by itself time-inhomogeneous offers the
advantage that the selection of Omori processes does not imply the need
of identifying how their non-stationarity emerges from an otherwise stationary
global behavior. In a version of our model suited for describing single,
long time series of returns \cite{stella,zamparo}, the necessity to consider random exogenous
factors influencing the market, leads us to switch-on at random
times some time-inhomogeneities formally similar to those characterizing the
model of the previous Section. This is achieved by setting $a_t=1$
concomitantly with these random events (See also
\cite{baldovin_0,baldovin,challet,andreoli}). 
In such a context it is not {\it a priori} clear whether or not the start
of an Omori process should imply putting $a_t=1$ in correspondence with
the time $t$ of the main shock. This difficulty is also accompanied by
the need of implementation of a more complicated 
calibration procedure \cite{zamparo} with respect to the one presented
here. 
 
As shown below, remarkable results of our analysis in this note
are: 
\begin{enumerate}
\item that the selected
processes are legitimately classified as Omori-like in the sense
that they can all be fitted by the Omori law; 
\item that the description one 
obtains for them based on the model presented in the previous Section
contains explicit dependencies on the intensities of the main shock
and on the aftershocks thresholds.
\end{enumerate}
This endows our approach to the Omori regimes of a predictive
potential which, if confirmed by further analysis, 
could be exploited by decision-makers under crisis conditions.

We select as Omori processes all those histories in
the above S\&P 500 ensemble for which the initial absolute return, $|r_0|$,
besides being smaller than $\sigma_{max}$, also exceeds a main shock
threshold $\sigma_m$.  
At variance with the analysis in 
Refs. \cite{lillo,lillo_1,selcuk,selcuk_1,weber,mu,petersen}, 
we consider,
in place of a single time series, groups of histories for which
$\sigma_m\leq|r_0|\leq\sigma_{max}$. 
As far as the aftershocks are concerned, we record for each of these
histories the elementary returns which exceed in absolute value an
aftershock threshold $\sigma_a$ and are below the main shock value
$|r_0|$: $\sigma_a\leq |r_i|\leq|r_0|$, for $i\geq 1$. 
The parameter $\sigma_a$ is an important one to be fixed
in any analysis of the Omori law.
Again, by imposing the aftershock magnitude to be smaller than that of
the main shock we reduce the influence of extreme events in our
limited dataset. 
We decided to search for main shocks occurring right at the beginning
of the daily time window described by our model for two main reasons.
In first place the ensemble average volatility on 10 minutes intervals
is maximal in the first interval. Secondly, a main shock 
occurring right at the beginning of the time window leaves the maximum 
possible time for the development of the subsequent Omori process.
While we will limit ourselves below to discuss such optimal case,
different choices are of course possible.

According to the above selection procedure of the Omori processes, the
cumulative number of aftershocks $N_{|r_0|}(t)$ after a main shock of
magnitude $|r_0|$ is given by
\begin{equation}
N_{|r_0|}(t)=
\mathbb E\left[
\sum_{i=1}^t
\mathbb 1_{\left(
\sigma_a\leq |R_i|\leq|R_0|
\right)}
\;|\;
|R_0|=|r_0|
\right],
\end{equation}
where $\mathbb 1_{\left(\sigma_a\leq |R_i|\leq|R_0|\right)}$ is the indicator
function, yielding $1$ if
$\sigma_a\leq |R_i|\leq|R_0|$ and zero otherwise.
Using Eq. (\ref{eq_student}), through a change of variable 
it is straightforward to show
\begin{eqnarray}
N_{|r_0|}(t)
&=&
\sum_{i=1}^t
\;2\;
\int_{\sigma_a}^{|r_0|} d r_i
\frac{p_{R_0,R_i}(r_0,r_i)}{p_{R_0}(r_0)}
\nonumber\\
&=&
\frac{2}{\sqrt{\pi}}
\;\frac{\Gamma\left(\frac{\alpha+2}{2}\right)}{\Gamma\left(\frac{\alpha+1}{2}\right)}
\;\sum_{i=1}^t
\int_{\frac{\sigma_a}{a_i\sqrt{\beta^2+r_0^2}}}^{\frac{|r_0|}{a_i\sqrt{\beta^2+r_0^2}}} d x
\;(1+x^2)^{-\frac{\alpha+2}{2}}.
\label{eq_omori_prediction}
\end{eqnarray}
If in the considered ensemble of histories there are $M$ realizations
$\left\{r_i^{(m)}\right\}_{m=1,2,\ldots,M}$ in
which we register a main shock, i.e.,
$\sigma_m\leq|r_0^{(m)}|\leq\sigma_{max}$, then the cumulative
number of aftershocks $N(t)$ is obtained through the sample average
\begin{equation}
N(t)=\frac{1}{M}\;\sum_{m=1}^MN_{\left|r_0^{(m)}\right|}(t),
\label{eq_omori_average}
\end{equation}
where we stress the fact that each $N_{\left|r_0^{(m)}\right|}(t)$ is
conditioned to the main shock magnitude $\left|r_0^{(m)}\right|$.
Notice that since with the available dataset the selected main shocks constitute a small
sample (see next Section), we use here the
sample average rather than the ensemble one to get the number of aftershocks
conditioned to $\sigma_m\leq|R_0|\leq\sigma_{max}$.

\begin{figure}[t]
%\sidecaption[t]
\begin{center}
\includegraphics[scale=.5]{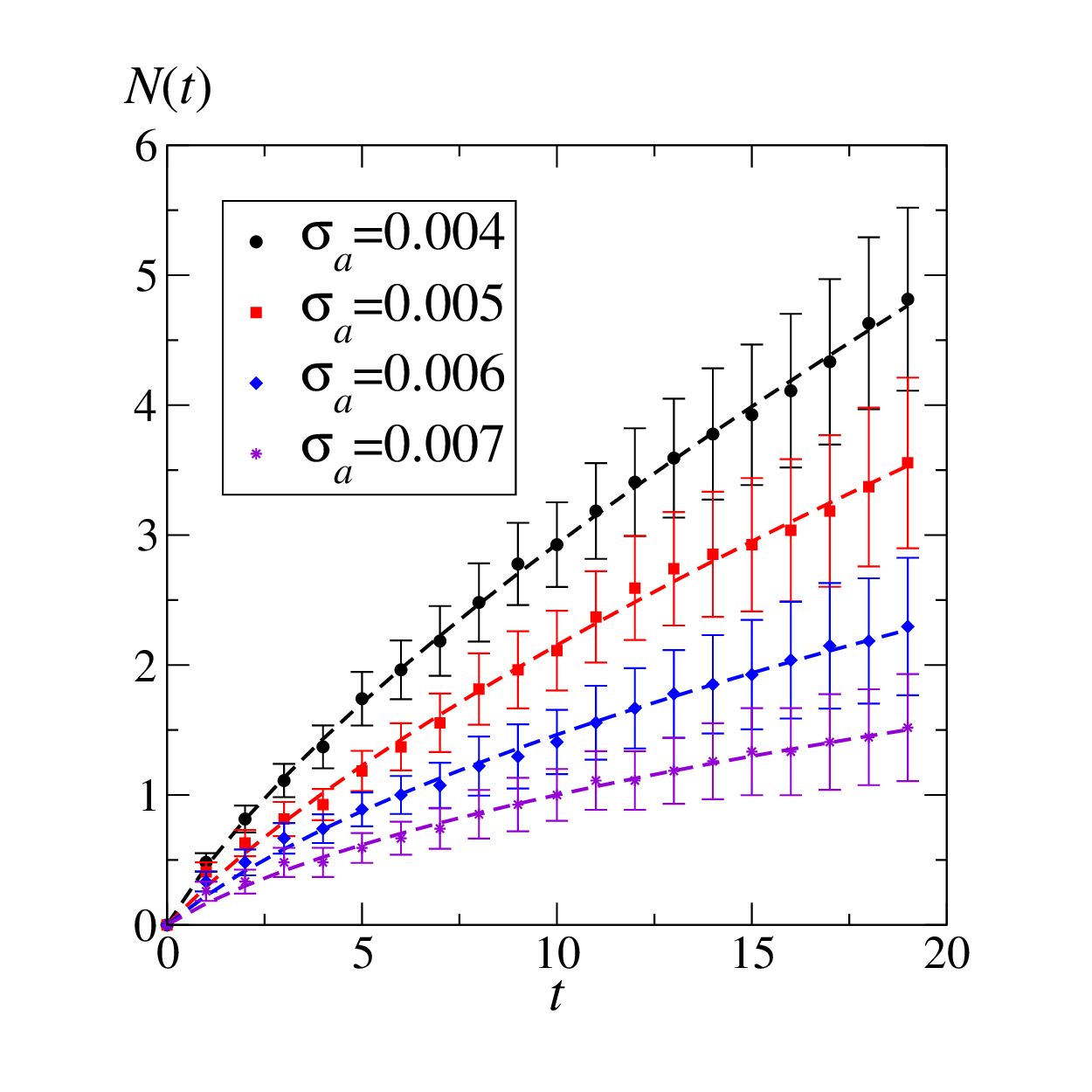}
\end{center}
\caption{
Fitting of the empirical aftershock at different thresholds $\sigma_a$
(points with error bars), with the Omori law in
Eq. (\ref{eq_omori_cumulative}) (dashed lines). Fitted parameters are
reported in Table \ref{tab_omori}.
}\label{fig_omori_1} 
\end{figure}

\begin{table}
\caption{Omori parameters in Eq. \ref{eq_omori_cumulative} fitted from
the empirical data.}
\label{tab:1}       % Give a unique label
%
% Follow this input for your own table layout
%
\begin{tabular}{p{2cm}|p{2.cm}p{2cm}p{2.cm}}
\hline\noalign{\smallskip}
$\sigma_a$ & $K$ & $p$ & $\tau$ \\ 
\noalign{\smallskip}\svhline\noalign{\smallskip}
$4\cdot10^{-3}\;$ & $\;0.44$ & $0.29$ & $0.52$ \\
$5\cdot10^{-3}\;$ & $\;0.40$ & $0.34$ & $2.00$ \\
$6\cdot10^{-3}\;$ & $\;0.35$ & $0.49$ & $2.00$ \\
$7\cdot10^{-3}\;$ & $\;0.28$ & $0.59$ & $2.07$ \\
\noalign{\smallskip}\hline\noalign{\smallskip}
\end{tabular}
\label{tab_omori}
\end{table}

\section{Comparison of the model predictions with the statistics
of aftershocks}
Our choice for the thresholds $\sigma_m$ and $\sigma_{max}$
is such that the absolute first returns for which 
$\sigma_m\leq|r_0^{(m)}|\leq\sigma_{max}$ are quite exceptional. 
They occur with $27/6283\simeq0.4\%$
frequency in our ensemble; Only 3 realizations have
$|r_0|>\sigma_{max}$ and are thus excluded. 
Accordingly, we analyze the averaged
$N(t)$ of aftershocks for these $M=27$ main shocks. 
A first point to clarify is whether the recorded rates
are well fitted by the Omori law in
Eq. (\ref{eq_omori_cumulative}). 
This is shown
in Fig. \ref{fig_omori_1}, where many sets of data for $N(t)$,
obtained with different aftershocks thresholds $\sigma_a$, 
are indeed fitted
by the Omori Eq. (\ref{eq_omori_cumulative}). 
In Table \ref{tab_omori} one also realizes that $K$, $\tau$ and $p$
need to be varied for each $\sigma_a$ in order to reach a satisfactory 
fit. In particular, by changing $K$ and $\tau$ it is even possible to
obtain reasonable fittings also with $p>1$ (See also \cite{mu}).  
This parameters variability makes it very difficult to use the
Omori law to predict the aftershock occurrence for a given main shock
magnitude and aftershock threshold.

\begin{figure}[t]
%\sidecaption[t]
\begin{center}
\includegraphics[scale=.5]{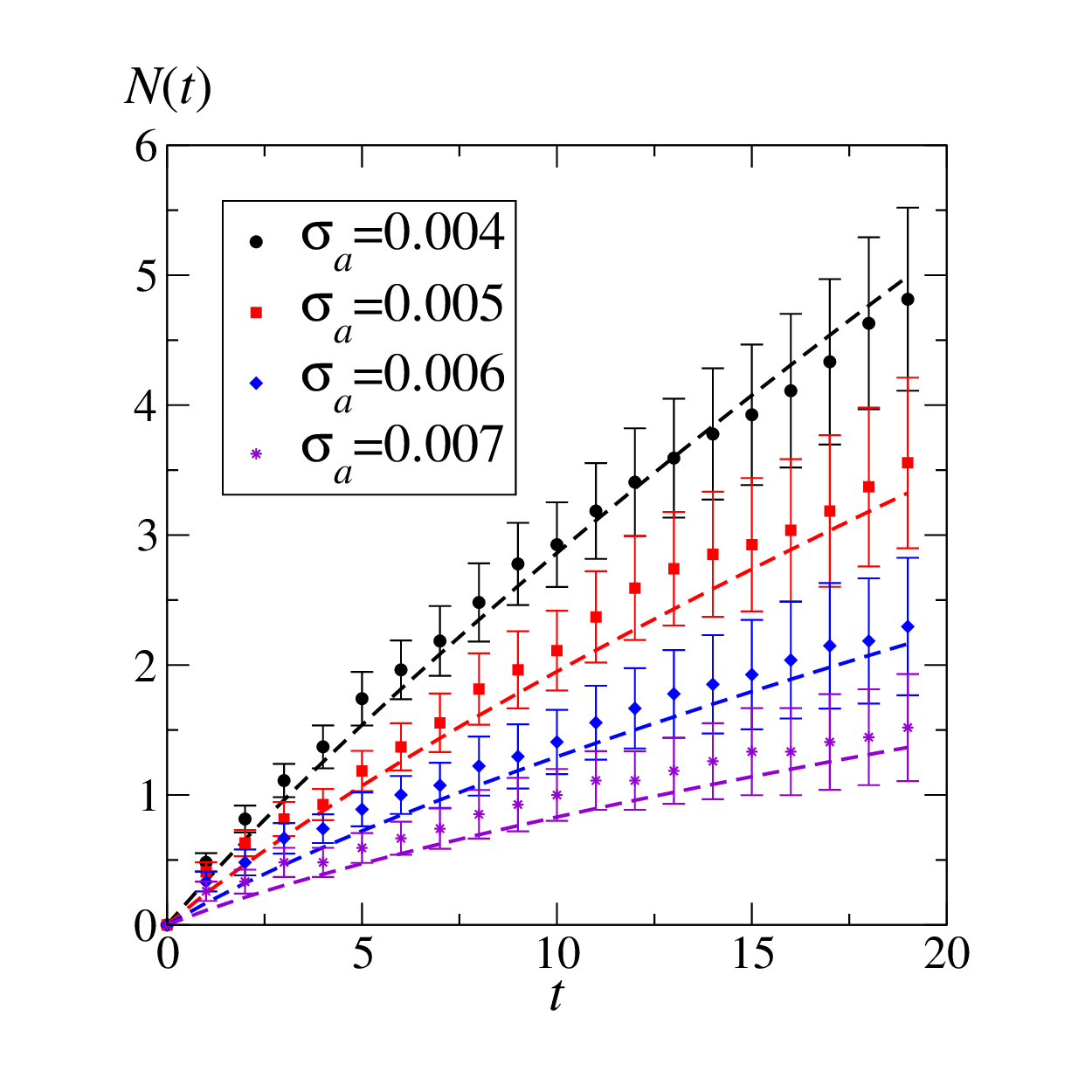}
\end{center}
\caption{
Comparison between the analytical model predictions for different
aftershock thresholds $\sigma_a$ (dashed lines) with the same empirical
S\&P data reported in Fig. \ref{fig_omori_1} (points with error bars).
}\label{fig_omori_2} 
\end{figure}

Model predictions on the same set of data fitted in Fig
\ref{fig_omori_1} are instead given in Fig. \ref{fig_omori_2}.
Dashed lines in Fig. \ref{fig_omori_2} are obtained on the basis
of Eqs. (\ref{eq_omori_prediction},\ref{eq_omori_average})
with the parameters $(\alpha,\beta,D)$ 
resulting from the calibration discussed
in Section \ref{sec_calibration}. 
The only difference among the curves is the value of the
aftershock threshold $\sigma_a$. 
The agreement of the analytical predictions with the data 
and the sensitivity of the curves to the variation of the aftershock
threshold are remarkable. 
This shows that our model potentially provides
a satisfactory and parameter-free description of Omori processes.

\section{Conclusions}
We have shown here in the case of the S\&P 500 index, that a model 
suited for the description of the high frequency market dynamics 
allows also to predict Omori regimes following exceptional extreme events.
Within the class of events considered,
the model specifies the dependence on the main shocks 
intensities and on the aftershocks threshold. As such, its description 
goes far beyond the limits of the Omori phenomenological law.

Besides providing a further validation of the model of
Refs. \cite{baldovin,baldovin_1}, 
the results presented here encourage to extend similar analysis
to cases in which the Omori processes are to be selected within
a process which is globally stationary. For the modeling
of these processes, our recipe \cite{zamparo} 
is that of switching-on at random
some non-stationarities ascribable to coefficients like the $a_t$
defined above. Global stationarity of the process on long time scales
is then guaranteed by the fact that empirical averages are
in this case made by considering time intervals sliding along
the single long history \cite{zamparo}. While it is conceivable that
in many cases main shocks are localized close to resets of the
time inhomogeneity, this is not true in general. Some attempts to
strictly identify main shocks with restarts of them inhomogeneity
in the model ($a_t=1$) already gave some preliminary agreement with the data. 
A more general discussion is however needed \cite{zamparo}.

\begin{acknowledgement}
This work is supported by 
``Fondazione Cassa di Risparmio di Padova e Rovigo'' within the 
2008-2009 ``Progetti di Eccellenza'' program. 
\end{acknowledgement}

%%%%%%%%%%%%%%%%%%%%%%%% referenc.tex %%%%%%%%%%%%%%%%%%%%%%%%%%%%%%
% sample references
% %
% Use this file as a template for your own input.
%
%%%%%%%%%%%%%%%%%%%%%%%% Springer-Verlag %%%%%%%%%%%%%%%%%%%%%%%%%%
%
% BibTeX users please use
% \bibliographystyle{}
% \bibliography{}
%

\end{document}